\documentclass[runningheads]{llncs}
\usepackage{graphicx}
\usepackage[hidelinks]{hyperref}
\usepackage[inline]{enumitem}

\usepackage{amsmath, amssymb}

\newcommand{\D}{\mathcal{D}}

\newcommand{\E}{\mathsf{E}}

\usepackage{xcolor}

\begin{document}
\title{Category Theory in Isabelle/HOL as a Basis for Meta-logical Investigation -- Preprint}
\titlerunning{Category Theory in Isabelle/HOL as a Basis for Meta-logical Investigation}
%
\author{Jonas Bayer\inst{1}\thanks{Corresponding authors, contributed equally}
\and
Alexey Gonus\inst{1,\star}
\and
Christoph Benzmüller\inst{2,1}
\and Dana S. Scott\inst{3}
}
\authorrunning{J. Bayer, A. Gonus, C. Benzmüller, and D. S. Scott}
%
\institute{Freie Universität Berlin, Germany 
\and Otto-Friedrich-Universität Bamberg, Germany
\and University of California \& Topos Institute, Berkeley, CA, USA }
%
\maketitle              
\begin{abstract} \sloppy
This paper presents meta-logical investigations based on category theory using the proof assistant Isabelle/HOL. We demonstrate the potential of a free logic based shallow semantic embedding of category theory by providing a formalization of the notion of elementary topoi. Additionally, we formalize symmetrical monoidal closed categories expressing the denotational semantic model of intuitionistic multiplicative linear logic. Next to these meta-logical-investigations, we contribute to building an Isabelle category theory library, with a focus on ease of use in the formalization beyond category theory itself. This work paves the way for future formalizations based on category theory and demonstrates the power of automated reasoning in investigating meta-logical questions.

\end{abstract}
%
%
%
%
%
%
\section{Introduction}
\label{sec:intro}

Category theory is a very abstract and general theory of mathematical structures~\cite{CftWM} that next to being used for organizing mathematical theories can also serve as an axiomatic basis of mathematics. It has a myriad of use cases in fields ranging from topology and algebra to the foundations of mathematics. 

A good understanding of categorical notions and methods can provide a mathematician with a generic framework to unify and describe concepts. The results obtained on a category theoretical level might later be applied to particular mathematical objects collected under a specific categorical setting. By this approach, many findings and ideas in one theory of mathematical structures can possibly be translated to the other. 

\paragraph{Formalizing Category Theory.}
Given its special standing in mathematics it is only natural to ask for a formalization of category theory, hoping that the benefits of the categorical perspective will carry over to formal mathematics. Concretely, one would not only wish to be able to formally use certain theorems from category theory but also utilize its power in the organization of formal mathematical libraries.  

Yet, the formalization of category theory poses significant challenges. Many experts consider first-order logic and the Zermelo-Fraenkel axioms of set theory as a suitable foundation for mathematics. However, e.g., topologists often work with the (large) category of topological spaces, which cannot be easily represented within this system of axioms, since large categories in standard formalizations within such a system are (obviously) not sets. It is thus not immediately clear how category theory should best be done formally. 

Next to this, significant challenges in the formalization of category theory stem from the double position that it carries in mathematics. Ideally, a formalization of category theory would not only lend itself to use cases in algebra but also enable meta-logical investigations that use categories as a foundation of mathematics. We refer to \cite{ernst2017,maddy2017,maddy2019} for a deeper discussion on category-theoretic versus set-theoretic foundations of mathematics.

\paragraph{Meta-logical investigations.} 
In this paper we intend to demonstrate the potential of a (shallow semantical embeddings based) formalization of category theory for the investigation of meta-logical questions. This will be done using the proof assistant Isabelle/HOL which is naturally suited for such work due to its strong support for automation.

Concretely, we give a formalization of the notion of elementary topoi that carry an important role in fundamental mathematics. We build up category theoretical concepts in order to eventually provide an elegant definition of a topos. Moreover, we formalize all necessary concepts allowing for future work that could implement the internal language of a topos.  

As a second meta-logical result, we present a formalization of linear logic (LL). Linear logic carries the idea of treating mathematical ``truths'' as information resources and has found a large number of theoretical and practical applications, ranging from computer science to linguistics. Categories come with the inherent property of representing denotational semantical models for different logics. We develop symmetrical monoidal closed categories that express the denotational semantical model of intuitionistic multiplicative linear logic (IMLL).

In addition, at an orthogonal, methodological level, we study the scalability of an approach to universal meta-logical reasoning \cite{ULRclasHOL}, that is based on shallow semantical embeddings of (layers of) object logics in classical higher-order logic, aka Church's simple type theory.

\paragraph{Further contributions.}
Next to our main focus on meta-logical investigation, we aim at contributing to the build up of a category theory library in Isabelle. At the time of writing, we are not aware of cases where the existing formalizations in the Isabelle Archive of Formal Proofs~\cite{Category-AFP,Category2-article,Category3-AFP,Category4-AFP} have been used for the purpose of verification in other fields of mathematics. Our work cannot (yet) compete with the extent of concepts verified in the aforementioned AFP entries. However, we pay a lot of attention to closely mimicking mathematical notation, cleanly organizing our theories, and giving examples of how to use the concepts we provide. Thereby, we hope to facilitate the use of category theoretical notions in future formalizations in other fields, too. Moreover, the paper is embedded in a larger project context, namely the exploration of the Benzmüller and Scott ~\cite{FreeLogCatTh} approach to the axiomatic modeling (in the tradition also of the early work of Saunders Mac Lane, although with more emphasis on Dana Scott, Freyd an Scedrov works) of category theory based on free-logic using the LogiKEy meta-logical/logico-pluralistic KR\&R~\cite{LogiKEY} methodology. It was important for us to study the scalability, advantages and disadvantages of this distinguished approach. In the end, we obtain a high degree of automation and a high level of abstraction. 

\section{Category Theory from a Free Logic perspective}
Our work exploits a shallow semantic embedding of free logic in Isabelle/HOL (or, more generally, in Church's simple type theory \cite{J43} aka. classical higher-order logic HOL), that is subsequently used for defining a notion of category theory, which then provides the basis for further formalization studies on top of it.

\subsection{Free Logic and its SSE into Isabelle/HOL}
\label{sec:FLinIsablle}
\paragraph*{Free Logic (FL)} is a logic that comes with less existential assumptions than its classical counterpart. Terms in free logic might denote so-called \textit{non-existent objects}, i.e., terms that refer to objects outside the domain of discourse~\cite{FreeLogicPl}. Existential and universal quantifiers are assumed to range over the \textit{existent terms}, i.e., those that denote objects within the domain of discourse. Such a logic is particularly interesting because it helps to reason about partiality. 

Therefore, free higher-order logic (FHOL) is ideal for an axiomatization of category theory since the composition of morphisms in a category is a partial function. In order to distinguish existent and non-existent objects we use the dual domain approach, i.e., we consider a domain $\D$ of all objects which has a subset $E$ of objects that are considered to be the existent objects. Alternatively, one could consider two disjoint sets for the non-existent and existent objects. We will follow the first approach as has been previously discussed by Cocchiarella~\cite{ddaFL} and also in the early work of Scott~\cite{scott1967existence}. The issue of properly defining free (higher-order) logic within the simple theory of types has been addressed in the works of Schütte~\cite{CHOLinFHOL1} and Farmer~\cite{CHOLinFHOL2} whose approach we follow with some modifications. It should also be noted that free logic could be implemented with exactly one undefined value (Benzmüller and Scott have also shown that before~\cite{AutoFreeLog}). Here we decided to go without this additional requirement (of having one undefined value) and to focus more on the ``existence'' part of the category theory without carrying about ``non-existing'' area.

\paragraph{Shallow semantical embedding.}
In order to reason formally and interactively within free logics without building a new theorem prover from scratch, a translation of logics is necessary. In Isabelle/HOL one can implement alike using a shallow semantical embedding (SSE), which is based on logic translation approaches as discussed by Gabbay, Nonnengart, Ohlbach and de Rijke~\cite{GabLDSys,Gabbay2SSE} for translating e.g. propositional modal logics to first-order logic. Exploiting the expressivity and compositionality of the simply typed $\lambda$-calculus in HOL, the SSE approach encodes such logic translations directly in HOL itself, which makes external translation mechanisms superfluous. This HOL-internal translation approach has been successfully extended for various quantified non-classicals logics and applied, under the name \emph{universal meta-logical reasoning}~\cite{ULRclasHOL}, amongst others, to encode free first-order logic~\cite{AutoFreeLog} in Isabelle/HOL.  This approach was then further extended to embed FHOL in HOL~\cite{FreeHOLPosit} which, in this paper, we will rely on to implement our higher-level categorical constructions. 

In SSE, the semantics of the language of interest, e.g. (positive) FHOL\footnote{Positive FHOL refers to a semantics for FHOL where formulas built from non-existing objects are allowed to be true ~\cite{FreeHOLPosit}.} is mapped to the corresponding syntax constructs of the target language. It may be viewed as a translation between the logics, where only semantical differences are targeted, for example, these could be the existential features of free semantics. The SSE approach showed itself as a readily available way for implementing the translation of a variety of nonclassical logics. It also enables the use of automation from the target system which is not as well supported with a deep semantical embedding.

A shallow semantic embedding is to be contrasted with a deep semantic embedding, in which the syntax of the target language is represented using an inductive data structure (e.g., following the BNF of the language) and the semantics of a formula is evaluated by recursively traversing the data structure. Shallow semantic embeddings, by contrast, define the syntactic elements of the target logic while reusing as much of the infrastructure of the meta-logic as possible; cf.~also \cite{J47}. In particular, the degree of proof automation that can be achieved is much better in the case of shallow semantic embeddings, since e.g. inductive proofs on the structure of the embedded logic are omitted.

\paragraph{Formalization in Isabelle/HOL.} Concretely, we represent the domain of objects $\D$ through a type $\alpha$ in Isabelle/HOL. The notion of existence is then given through a predicate $\E : \alpha \to \mathrm{bool}$. Therefore, every function will be total when viewed on the level of Isabelle terms. However, from the free logic perspective non-partial maps can still be observed as such since they are modelled as functions that map some objects to ``non-existent'' objects outside $E$. 

At the level of free logic, one can immediately define several notions of equality, which are used in the definition of categories and reappear in the course of the development of the formalizations.

\begin{definition}[Equalities with Existence]
\label{def:equalities}
Given $x, y \in D$ define three notions of equality as follows:
\begin{enumerate} 
\item We write $x\simeq y$ if and only if $x=y\land \E x \land \E y$ (Existing Identity). 
\item We also write $x\cong y$ if and only if $(\E x \lor \E y) \longrightarrow x=y$ (Kleene Equality). 
\item Finally, we write $x \geq y$ precisely when $\E x \longrightarrow x=y$ (Directed equality).
\end{enumerate}
\end{definition}

\subsection{Formalization of Axiomatic Category Theory}
\label{sec:ACT}
In Isabelle/HOL, concepts from category theory have been formalized as early as 2005~\cite{Category-AFP}. The original formalization could be improved and extended significantly as indicated by subsequent research~\cite{Category2-article,Category3-AFP,Category4-AFP}. 

In addition to these, Benzmüller and Scott presented an alternative approach for formalizing category theory in Isabelle/HOL~\cite{FreeLogCatTh}, which is based on an axiom system in free logic originally proposed by Scott~\cite{Scott1979}. This work models on one-sorted categories, i.e., it only refers to morphisms without mentioning objects. This approach was first expanded upon by Tiemens who defined inverse categories in order to generalize so-called modeloids~\cite{modeloids}.

\paragraph{Categories.} 
Our formalization of categories follows the approach by Benzmüller and Scott~\cite{FreeLogCatTh} with slight modifications\footnote{It should be noted that all categories considered in this paper are one-sorted categories and, therefore, all free variables appearing in the definitions refer to ``morphisms'', although, they might be seen as ``objects'' in a usual sense when they satisfy specific (identity) predicates.}. Firstly, when declaring the categorical notions of domain, codomain and composition, polymorphic types are employed which allow the use of higher-level constructions later. Secondly, an additional axiom is added that states the existence of a ``non-existent object''. This means that $\D$ is required to be a strict superset of $E$ which is advantageous in the definition of certain concepts as it enables referring to an explicit non-existent element. On the implementation side, we represent categories through an Isabelle locale. 

\paragraph{Functors.}
Functors are the morphisms between categories. The next definition is slightly adapted from Freyd and Scedrov's textbook~\cite{CatsAllegories}:\footnote{The first and second axioms result from the totality of functions in Isabelle/HOL and are used for the separation of existing and non-existing morphisms. This part of the paper, where we have to deal with the proper preservation of existence, has been one of the main difficulties in the subsequent formulation of the concepts.}

\begin{definition}[Functor] \sloppy 
\label{def:func}
A {\it functor} {\bf F} between two categories $\mathcal{C}$ and $\mathcal{D}$ is a function $\textbf{F}:\mathcal{C}\longrightarrow\mathcal{D}$ which satisfies the following axioms:
\begin{enumerate*}[label=(\arabic*)]
\item  $\textbf{E}x\longrightarrow \textbf{E}(F(x))$, 
\item $\neg\textbf{E}x\longrightarrow \neg\textbf{E}(F(x))$, 
\item $\textbf{F}(dom_{\mathcal{C}}(x)) \cong dom_{\mathcal{D}}(\textbf{F}(x))$, \item  $\textbf{F}(cod_{\mathcal{C}}(x)) \cong cod_{\mathcal{D}}(\textbf{F}(x))$, \item $\textbf{F}(x\cdot_{\mathcal{C}}y) \geq \textbf{F}(x)\cdot_{\mathcal{D}}\textbf{F}(y)$.
\end{enumerate*}
\end{definition} 

\paragraph{Natural Transformations.}
Morphisms between functors are called natural transformations. There can be two different formulations, resp.~formalizations, of this notion, and both are used in our work. The first definition is taken from \cite{NatTransformNlab} (the one which is directly based on the idea of one-sorted categories and is heavily exploited in this formalization) and modified according to the equalities that were introduced earlier:

\begin{definition}[Natural Transformation]\sloppy
\label{def:nat1}
A {\it natural transformation} $\eta$ between the functors {\bf F}$:\mathcal{C}\longrightarrow\mathcal{D}$ and {\bf G}$:\mathcal{C}\longrightarrow\mathcal{D}$ is a function $\eta:\mathcal{C}\longrightarrow\mathcal{D}$ such that: 
\begin{enumerate*}[label=(\arabic*)]
\item
$\textbf{E}x\longrightarrow \textbf{E}(\eta(x))$, \item
$\neg\textbf{E}x\longrightarrow \neg\textbf{E}(\eta(x))$, \item
$dom_{\mathcal{D}}(\eta(x)) \cong dom_{\mathcal{D}}(\textbf{F}(x))$, \item
$cod_{\mathcal{D}}(\eta(x)) \cong cod_{\mathcal{D}}(\textbf{F}(x))$, \item
$\textbf{E}x\cdot y\longrightarrow\eta(x)\cdot_{\mathcal{D}}\textbf{F}(y) \simeq \textbf{G}(x)\cdot_{\mathcal{D}}\eta(y)$.
\end{enumerate*}
\end{definition}

The monoidal category characterization is partly based on the notion of \emph{inverse natural transformation}, which is more naturally described with the following second definition:

\begin{definition}[Natural Transformation through Identities]
\label{def:nat2}
A {\it natural transformation} $\eta:\textbf{F}\Rightarrow\textbf{G}$ between the functors {\bf F}$:\mathcal{C}\longrightarrow\mathcal{D}$ and {\bf G}$:\mathcal{C}\longrightarrow\mathcal{D}$ assigns to every object $A$ a morphism $\eta(A):\textbf{F}(A)\longrightarrow\textbf{G}(A)$, such that for any morphism $x:A\longrightarrow B$ in $\mathcal{C}$ $\textbf{G}(x)\cdot_{\mathcal{D}}\eta(dom_{\mathcal{C}}(x)) \cong \eta(cod_{\mathcal{C}}(x))\cdot_{\mathcal{D}}\textbf{F}(x)$.
\end{definition}

This second definition can be (and was) extended to the former via a specification of how this function operates in the more general case for all morphisms, i.e. for $x:A\longrightarrow B$ we have $\eta(x) = \textbf{G}(x)\cdot_{\mathcal{D}}\eta(dom_{\mathcal{C}}(x))$. Starting with these definitions we proceed further to define {\it natural isomorphisms} and {\it inverse natural transformations}. The later concept might be an example of a module system advantage as it is easily defined as a locale built on top of the natural isomorphism locale with the specification of the inverse mapping,. Invoking Isabelle's \texttt{unfold\_locales} method during proving then allows to split the conditions into simpler parts.

\section{Formalization of Elementary Topoi}
In order to perform the intended meta-logical investigations it is necessary to define additional structures on top of categories. In particular, we formalize notions like categories with binary (co)products, exponential categories and cartesian closed categories. Our implementation makes heavy use of Isabelle's locales that allow to elegantly model the layered character of these definitions. To validate the correctness of the implementation we also formalize certain examples including the category of categories and the category of sets. The Isabelle code of all concepts that will be presented in this section can be found in a gitlab repository.\footnote{\url{https://permalink.jonasbayer.de/bachelorthesis}} Our formalization of elementary topoi generally follows the book ``Elementary categories, elementary topoi'' by Colin McLarty~\cite{ml_92}.

\paragraph{Formalization of elementary notions.} To build up the necessary constructions, we first formalize various elementary structures that can be defined within a category. This includes initial and final objects, (co)products, equalizers, a generic implementation of limits, monomorphisms and epimorphisms, and pullbacks. Not only do we define these notions, but we also formalize the elementary equivalences and relations between them. Each of these categorical structures received their own Isabelle theory in order to increase clarity, with definitions made in a multi-layered style and custom notation introduced for ease of use.

\begin{figure}[h]
\centering
\includegraphics[width=\textwidth]{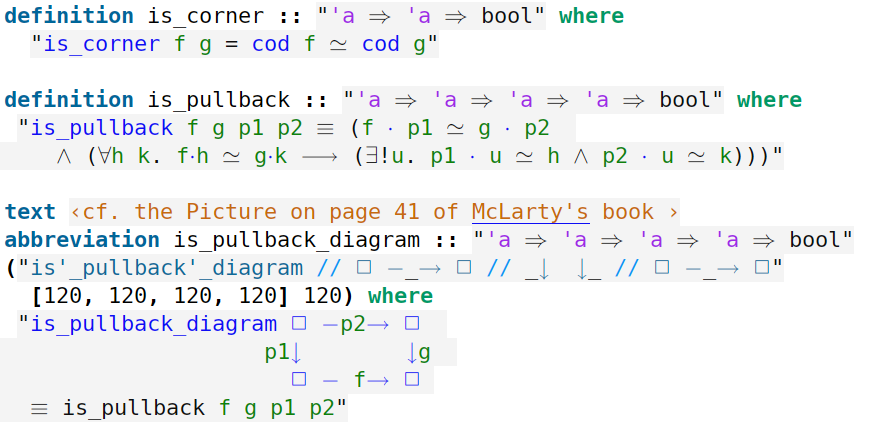}
\caption{The implementation of pullbacks in Isabelle}
\label{fig:pullback}
\end{figure}

As an example, consider the beginning of the theory on pullbacks shown in Fig.~\ref{fig:pullback}. There, we first introduce the preliminary notions of \texttt{is\_corner} and \texttt{is\_pullback} before defining pullback diagrams. The latter is given in custom syntax that allows a presentation in diagram format. Although such syntax can become cluttered during proving and therefore is not the most useful representation in that context, it still allows for a very readable presentation of results within the theorem prover. This approach is continued throughout the formalization, further examples include product diagrams or even commutative diagrams with multiple squares. 

Next to these constructions, we also give an implementation of several basic categories. Most notably, this includes the category of sets, the category of categories and the poset category. All these examples are formalized following the same scheme of first defining a custom type which will correspond to the type $\alpha$ in the category definition. Consequently, the existence predicate $\E$ followed by definitions for (co)domain and composition can be declared. In all three cases, the instance proof can essentially be handled by Isabelle's automatic tools. When using Isabelle2021 (February 2021), which is the version this development started with, occasional help is only necessary for proving associativity of composition. In Isabelle2022 even this part can be tackled by automated tools within less than 5 seconds when employed on an average personal computer. 

\paragraph{Categories with additional structure.} Refering back to the elementary structures defined previously, we implement categories that have additional structures. We start with categories that have binary products and/or binary coproducts. To validate these definitions, it is also formalized that the poset category has binary products and coproducts when the poset is a lattice. In this case, products correspond to meets and coproducts correspond to joins. 

Having defined binary products one can continue to declare exponential objects and exponential evaluation maps in a category which we collect into an Isabelle record. Exponentials are then used to specify cartesian closed categories after having defined cartesian categories. For the precise mathematical definitions, we refer the reader to McLarty~\cite{ml_92} whose presentation we follow closely.

\paragraph{Formalization of topoi.} With all these preliminary notions at hand, the definition of a topos can finally be formalized: 

\begin{figure}[h]
\centering
\includegraphics[width=\textwidth]{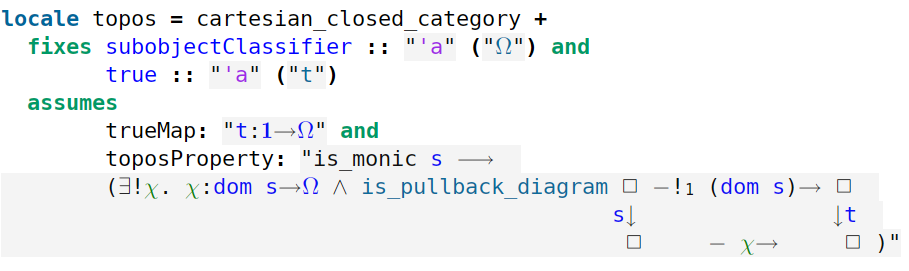}
\caption{The implementation of an elementary topos in Isabelle}
\label{fig:topos}
\end{figure}

A topos is a special case of a cartesian closed category that also has a so-called subobject classifier and a designated object $t$ representing the boolean value true. Here we make use of the aforementioned custom syntax to present the pullback diagram in a very intuitive form. Moreover, we formalize that in a topos all monomorphisms are equalizers, which does not hold in arbitrary categories. 

To conclude, the formalization we give provides a basis for for further meta-logical investigations related to category theory. In particular, an interesting continuation of this work would be the implementation of the internal language of an elementary topos including topos axioms.   

\section{Formalization of the Categorical model of IMLL}
\label{sec:IMLLform}

Another direction that has been explored in the proposed approach to categorical formalizations is the one laying down a translation layer between the HOL/FHOL and \emph{intuitionistic multiplicative linear logic} (IMLL) semantics. The motivation behind the IMLL formalization (or its semantics, to be more precise) was to investigate practical capabilities of Isabelle/HOL in the field of meta-logical questions, i.e., to derive tools that would formulate IMLL theorems and proofs through its encoded semantical language that talk about morphisms. It is important to emphasize the word ``practical'' in this context, since it is of course generally possible to model a wide range of constructions with the help of a proof assistant such as Isabelle/HOL.  However, it is not a priori clear if this can be done with a reasonable effort with respect to the logic ``translations'' as exploited in our SSE approach. In this paper we have chosen (a fragment of) linear logic as our object of study, since it has a plethora of applications in various domains.  The Isabelle implementation and certification of all constructions presented in this section (which make use of the ``locales" feature of Isabelle/HOL) can be found in a GitHub repository.\footnote{\url{https://github.com/HaskDev0/Linear-Logic-Cat-semantics}}

\subsection{IMLL and its Categorical Model}
\label{subsec:imll_cat_model}

{\it Linear Logic} (LL) was proposed by Girard in 1987 \cite{LLor} as a refinement of classical as well as intuitionistic logic. Within LL the two different notions of conjunction and disjunction are introduced together, i.e., {\it multiplicative} and {\it additive} types of connectives. They behave distinctly in derivations (through inference rules), but their intrinsic difference can be revealed with a computational interpretation of proofs  \cite{IntToLL}. Within this work we deal with a particular fragment of LL, which is {\it intuitionistic multiplicative linear logic} (IMLL), by restricting the syntax and adapting inference rules.  

The categorical denotational model of IMLL is built around the notion of a {\it monoidal category}.\footnote{For more information, see Mellies \cite{CSLL}. It provides a thorough description of the topic as well as lists all the inference rules for IMLL that are formalized and proved within our framework.} The key part of a semantical translation is a function $[\![ \cdot ]\!]$ that assigns some categorical construct to a proof-theoretical piece. Moreover, the construction of categories using only one sort of objects, i.e., morphisms, determines our main interest in modeling proofs (rather than formulas) with categorical denotational semantics. In order to briefly present the general idea, one assigns a morphism $f:[\![ A ]\!]\longrightarrow [\![ B ]\!]$ to a proof of a sequent $A\vdash B$.\footnote{All the translations of proofs are carried out under proof-theoretic normalization invariance.} Here, $[\![ A ]\!]$ means an identity morphism. The final goal within this part of the work is to translate all the inference rules we have in IMLL to theorems about morphisms in a special category, i.e., {\it symmetrical monoidal closed category} (SMCC), and to show  practical feasibility of these meta-logical translations. 

\subsection{Isabelle/HOL formalization of the IMLL Categorical Model}
\label{subsec:imll_form}

To fully represent IMLL connectives, such as linear intuitionistic implication and multiplicative conjunction, we need to incorporate their corresponding inference rules into our categorical setting. These rules describe the connectives' behavior as well as their structural and non-logical inference rules. To do so, we must make use of monoidal categories, braidings and symmetries, and closed structures.

The shift towards monoidal categories for modeling IMLL semantics is crucial due to the necessity to move away from diagonal maps in categories which allow the duplication of the objects, or, equivalently, resources \cite{ItCaCL}.\footnote{The definitions that are needed to describe the model of IMLL are taken from Mellies \cite{CSLL} and adopted to our framework.} Moreover, the \emph{locales} environments of Isabelle/HOL are used to develop formalizations of translated IMLL concepts for their convenience of presentation and of working with algebraic theories.

\begin{definition}[Monoidal Category] \sloppy
\label{def:mc}
    A {\it monoidal category} $\mathcal{C}$ is a category $\mathcal{C}$ equipped with: 
\begin{enumerate*}[label=(\arabic*)] 
    \item a bifunctor $\otimes:\mathcal{C}\times\mathcal{C}\longrightarrow\mathcal{C}$,
    \item a natural isomorphism\footnote{A family of natural isomorphisms.} $\alpha_{A,B,C}:(A\otimes B)\otimes C \longrightarrow A\otimes(B\otimes C)$, 
    \item a special identity morphism, or object in our meaning, {\bf e}, which is a unit, and \item two natural isomorphisms $\textbf{e}\otimes A \longrightarrow A$ and $\rho_A:A\otimes \textbf{e} \longrightarrow A$, which 
    satisfy the triangular identity 
    $(\mathcal{C}.Id\ A) \land (\mathcal{C}.Id\ B) \longrightarrow (A\otimes\lambda_B)\cdot\alpha_{A,\textbf{e},B}\simeq \rho_A\otimes B$
    and the pentagonal identity 
    $(\mathcal{C}.Id\ A) \land (\mathcal{C}.Id\ B) \land (\mathcal{C}.Id\ C) \land (\mathcal{C}.Id\ D)\longrightarrow
    \allowbreak (A\otimes\alpha_{B,C,D})\cdot(\alpha_{A,B\otimes C,D}\cdot\alpha_{A,B,C}\otimes D) \simeq \alpha_{A,B,C\otimes D}\cdot\alpha_{A\otimes B,C,D}$.\footnote{$\mathcal{C}.Id$ denotes an identity morphism predicate.} 
\end{enumerate*}    
\end{definition}

Here it is essential to understand that the natural isomorphism $\alpha$ acts between functors $(\bullet\otimes\bullet)\otimes\bullet:\mathcal{C}\times\mathcal{C}\times\mathcal{C}\longrightarrow\mathcal{C}$ and $\bullet\otimes(\bullet\otimes\bullet):\mathcal{C}\times\mathcal{C}\times\mathcal{C}\longrightarrow\mathcal{C}$ of domain $\mathcal{C}\times\mathcal{C}\times\mathcal{C}$ and codomains $\mathcal{C}$. There exist similar interpretations for the other two isomorphisms occurring in the definition.

\begin{definition}[Braided Monoidal Category]\sloppy
\label{def:braid_mc}
A {\it braided monoidal category} is a monoidal category $\mathcal{C}$ equipped with a {\it braiding}, i.e. a natural isomorphism $\gamma_{A,B}:A\otimes B \longrightarrow B\otimes A$,
making two hexagonal axioms hold: 
\begin{enumerate*}[label=(\arabic*)] 
\item $(\mathcal{C}.Id\ A) \land (\mathcal{C}.Id\ B) \land (\mathcal{C}.Id\ C) \longrightarrow \alpha_{B,C,A}\cdot(\gamma_{A,B\otimes C}\cdot\alpha_{A,B,C}) \simeq \allowbreak (B\otimes\gamma_{A,C})\cdot(\alpha_{B,A,C}\cdot(\gamma_{A,B}\otimes C))$,  and 
\item $(\mathcal{C}.Id\ A) \land (\mathcal{C}.Id\ B) \land (\mathcal{C}.Id\ C) \allowbreak \longrightarrow \allowbreak
\alpha^{-1}_{C,A,B}\cdot(\gamma_{A\otimes B,C}\cdot\alpha^{-1}_{A,B,C}) \simeq (\gamma_{A,C}\otimes B)\cdot(\alpha^{-1}_{A,C,B}\cdot(A\otimes\gamma_{B,C}))$.
\end{enumerate*}
\end{definition}

\begin{definition}[Symmetric Monoidal Category]
\label{def:sym_mc}
    A {\it symmetric monoidal category} is a braided monoidal category $\mathcal{C}$, whose braiding is a {\it symmetry}, i.e. $(\mathcal{C}.Id\ A) \land (\mathcal{C}.Id\ B) \longrightarrow\gamma_{A,B} \simeq \gamma^{-1}_{B,A}$.
\end{definition}

These notions already allow one to reason about the multiplicative conjunction and exchange inference rules in IMLL, and the final step is to properly describe
{\it closed structures}. There are three equivalent definitions of these terms, and the way to define one chooses depends on the goals and the frame one works in due to the ease of use in some cases. We will give two variations here and we will provide some comments on their connection. The first one was chosen since it will directly be used in further developments of IMLL semantics, and the second one will deliver a clearer view of the notion and the understanding of why we introduced an evaluation morphism $eval$ later (see Def. \ref{def:left_cs_isa2}) to describe the translation of formulas to categorical constructs.

\begin{definition}[Left Closed Monoidal Category]
\label{def:left_cs}
A {\it left monoidal closed category} is a monoidal category $\mathcal{C}$ endowed with a {\it left closed structure}, i.e. with a data of: 
\begin{enumerate*}[label=(\arabic*)] 
\item a {\it bifunctor} $\multimap:\mathcal{C}^{op}\times\mathcal{C}\longrightarrow\mathcal{C}$, and
\item a {\it bijection} $\mathcal{C}(A\otimes B,C) \cong_b \allowbreak \mathcal{C}(B,A\multimap C)$, which is natural in $A,B$ and $C$.\footnote{The sign $\cong_b$ here means the bijection between the sets and we introduced it in order to distinguish it from the Kleene duality.}
\end{enumerate*}
\end{definition}

The second point in the definition should be understood as the natural transformation between the functors $\mathcal{C}(\_\otimes_{ex}\_,\_)$ and $\mathcal{C}(\_,\_\multimap\_)$ which act between categories $\mathcal{C}^{op}\times\mathcal{C}^{op}\times\mathcal{C}$ and $Set$. Therefore, the encoding of $Hom$-functors should have been encoded first, which would have added additional layer of complexity. Instead, we decided to translate it to the language of sets directly with the help of two function $\Phi$ and $\Psi$ acting as inverses of each other and having more inputs than just one, namely, the morphism itself. The functions, thus, have the following types: 

\begin{center}
    $\Phi::'a\Rightarrow'a\Rightarrow'a\Rightarrow'a\  \text{and}\  \Psi::'a\Rightarrow'a\Rightarrow'a\Rightarrow'a.$
\end{center}

The reason we are using two functions describing the bijection lies in the fact that in the Isabelle/HOL system functions are total, but in our framework we are working with the so-called existent morphisms which constitute only a subdomain of some type. 
Moreover, two additional arguments need to be specified when we want to apply the mentioned bijection, in order to know the exact structure of the domain of the input morphism for the bijection going right in the Def.~\ref{def:left_cs}~(2) and the same for the codomain of the input morphism for the bijection going left. 
In other words, given some morphism $f$, and a necessity to check whether the domain of $f$ is exactly $A\otimes B$, we cannot merely apply the $dom$ function to find the hidden parts. Therefore, this additional information recovers the missing data. The same reasoning clarifies the equivalent features of $\Psi$. This encoding process also reveals the actual compressed information that is needed to talk about IMLL semantical constructs within Isabelle/HOL. Therefore, formalization of this definition in Isabelle/HOL looks as follows:

\begin{definition}[Left Closed Monoidal Category in Isabelle/HOL]
\label{def:left_cs_isa}
A {\it left monoidal closed category} is a monoidal category $\mathcal{C}$ endowed with a {\it bifunctor} $\multimap:\mathcal{C}^{op}\times\mathcal{C}\longrightarrow\mathcal{C}$ and functions $\Phi$ and $\Psi$, that satisfy: \vspace{-.5em}
\begin{enumerate}[label=(\arabic*)] 
\item $(f:A\otimes B\rightarrow C) \land (\mathcal{C}.IdEx\ A) \land (\mathcal{C}.IdEx\ B)\longrightarrow \Phi(A,B,f):B\rightarrow(A\multimap C)$, 
\item $(g:B\rightarrow A\multimap C) \land (\mathcal{C}.IdEx\ A) \land (\mathcal{C}.IdEx\ B)\longrightarrow \Psi(A,C,g):A\otimes B\rightarrow C$, 
\item $(f:A\otimes B\rightarrow C) \land (\mathcal{C}.IdEx\ A) \land (\mathcal{C}.IdEx\ B)\longrightarrow \Psi(A,C,\Phi(A,B,f)) \simeq f$,
\item $(g:B\rightarrow A\multimap C) \land (\mathcal{C}.IdEx\ A) \land (\mathcal{C}.IdEx\ B)\longrightarrow \Phi(A,B,\Psi(A,C,g))\simeq g$, and the requirement for $\Psi$ for being natural in $f$:
\item $(f:A\otimes B\rightarrow C) \land (\mathcal{C}.IdEx\ A) \land (\mathcal{C}.IdEx\ B) \land (cod\ x = A) \land (cod\ y = B) \land (dom\ z = C)\longrightarrow \Phi(dom\ x, dom\ y, z\cdot(f\cdot(x\otimes y)))\simeq(z\multimap y)\cdot(\Phi(A,B,f)\cdot y)$.
\end{enumerate}
\end{definition}
We have proved (in our formalization) that this function $\Psi$ is a natural transformation.
As was pointed out above, the other definition of {\it left closed structure} is the following: 

\begin{definition}[Left Closed Structure]
\label{def:left_cs_isa2}
A {\it left closed structure} in a monoidal category $\mathcal{C}$ is composed of: 
\begin{enumerate*}[label=(\arabic*)] \item an {\it identity morphism} $A\multimap B$, and 
\item a {\it left evaluation morphism} $eval_{A,B}:A\otimes(A\multimap B)\longrightarrow B$, 
for every identity morphisms $A$ and $B$. The $eval_{A,B}$ morphism satisfies the following universal property: \par $\forall f:A\otimes X\longrightarrow B.(\exists!h:X\longrightarrow A\multimap B)\land(f\simeq eval_{A,B}\cdot (A\otimes h)).\footnote{Here, we exploited a new abbreviation $\exists!h.P(h)$ for unique existence, which means $\exists h.(P(h) \land (\forall t.P(t) \to t=h))$.}$
\end{enumerate*}
\end{definition}

The evaluation morphism \emph{eval} is a translation of the elimination rule for $\multimap$ if the proof system is designed with it instead of the \emph{left} rule for $\multimap$. The former definition entails the latter for a left closed structure (this is formalized as a theorem in Isabelle/HOL). In the same way, with slight modifications, one may define the {\it right closed structure}.




At this point we are able to describe the categorical model of IMLL.

\begin{definition}[Symmetric Monoidal Closed Category]\sloppy
\label{def:smcc}
A {\it symmetric monoidal closed category} (SMCC) is a monoidal category $\mathcal{C}$ equipped with a {\it symmetry} $\gamma$ and a {\it left closed structure} $\multimap$.
\end{definition}

We proved in our formalization that in any such monoidal category there is also a {\it right closed structure} as well, which confirms theoretical expectations. To demonstrate this, define the functions $\bullet\multimap_r\bullet = \bullet\multimap\bullet$, $\Phi_r(f) = \Phi(f\cdot\gamma_{A,B})$ and $\Psi_r(g)= \Psi(g)\cdot\gamma_{B,A}$ for morphisms $f:B\otimes A\longrightarrow C$ and $g:B\longrightarrow A\multimap C$, which would exhibit the desired properties.\footnote{That is exactly the place where we use in the encoding those additional first two parameters of $\Phi$ and $\Psi$ to specify the morphisms.} 
Provided that we described through locales tensor product \texttt{T}, left closed structure \texttt{Impl}, natural isomorphisms $\alpha, \mu, \rho$, natural isomorphisms between \emph{Hom-}functors via $\Phi, \Psi$ and a symmetry $\gamma$, the corresponding encoding of a SMCC inside Isabelle/HOL would look like:

\begin{figure}[h]
\centering
\includegraphics[width=0.8\textwidth]{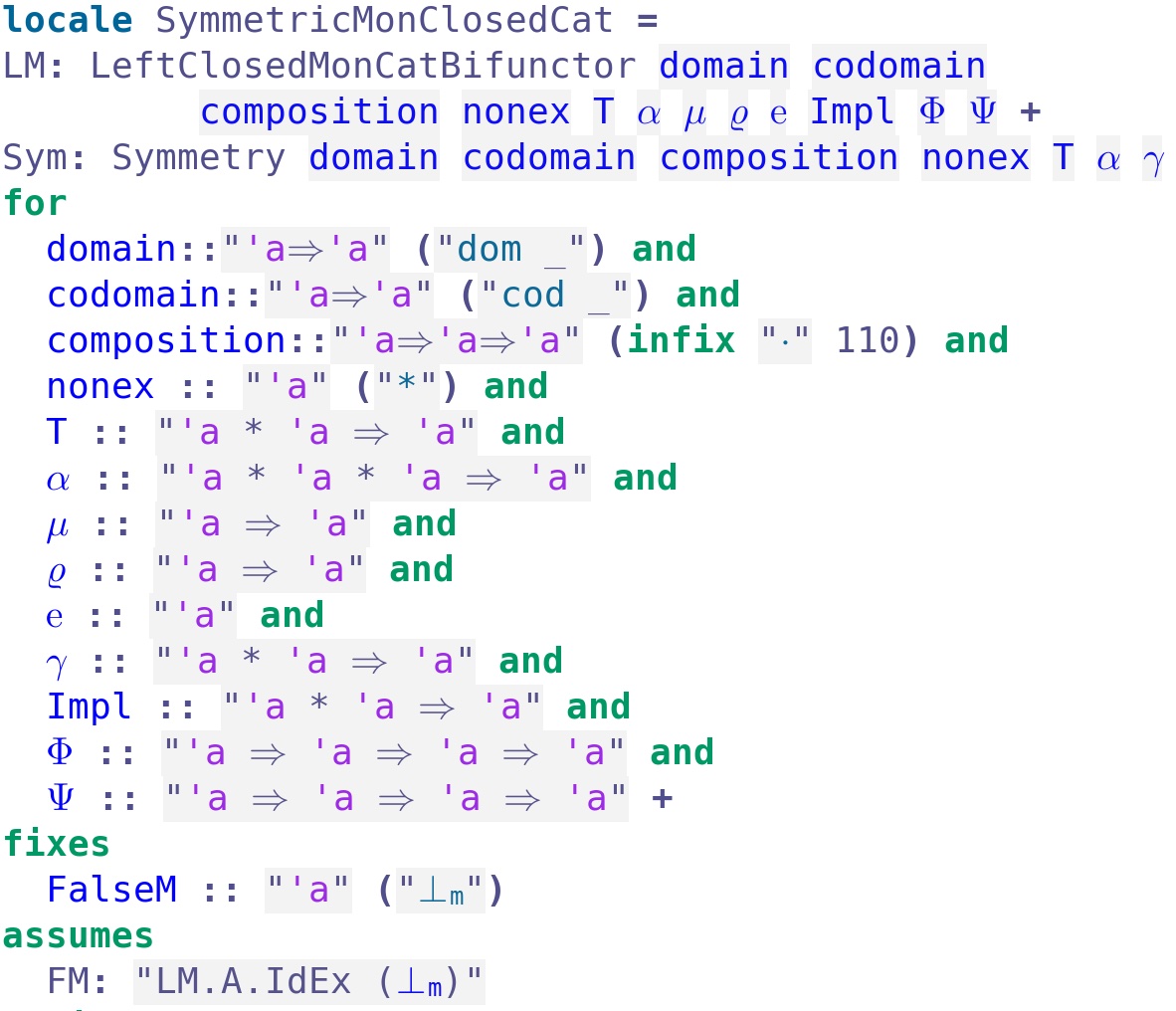}
\caption{The implementation of a symmetric monoidal closed category}
\label{fig:smcc}
\end{figure}


At this point we are ready to give the formalization of IMLL formulas in SMCC. For that, we start with the discussion of the well-known principle in intuitionistic logic, which in particular holds in IMLL, that every formula $A$ implies its double negation $\neg\neg A$. For this purpose there should be some formula $\bot$, 
which helps to define the negation of a formula as $A\multimap\bot$ \cite{CSLL}. Thus, the mentioned principle translates to the fact in SMCC as:

\begin{proposition}[Existence of Double Negation Morphism]
\label{prop:double_neg_in_smcc}
There is always a morphism $\delta_A:A\longrightarrow(A\multimap\bot)\multimap\bot$ for every identity $A$ in any SMCC.
\end{proposition}

Note that this is not a new result, but a proposition that we used to test our encoding with and which we certified within the Isabelle/HOL SSE framework.

It is worth mentioning that in a sequent style proof there is no difference  how we derive the formula $(A_2\multimap\bot)\multimap\bot$ given a formula $A_1$ and a derivation $\pi$ of $A_1\vdash A_2$. In other words, we could have derived to it via double negating $A_1$ and then applying the derivation $\pi$ translated for double negation, or we double negate $A_2$. In SMCC this fact corresponds to: 

\begin{proposition}[Double Negation as a Natural Isomorphism]
\label{prop:double_neg_as_nat_is}
In every SMCC the constructed morphism $\delta_A$ is, in fact, a natural transformation.
\end{proposition}

Propositions \ref{prop:double_neg_in_smcc}, \ref{prop:double_neg_as_nat_is} already quite firmly indicate the possibility of applying the chosen categorical axiomatization and formalization approach to logical questions expressed in a denotational (categorical) framework. As long as everything above is formalized in Isabelle/HOL, it is now possible to translate all the inference rules of IMLL as theorems about the morphisms in SMCC. For this step one has to be slightly creative in terms of finding the most appropriate translations, i.e., relying only on automated tools for this step would be a very naive way, leading to an explosion of ``search space'', while human interference with "reasonable" assumptions and ideas solves the problem of finding suitable morphisms that represent specific IMLL sequents. The task was successfully accomplished, as can be seen in the full encoding. This provides some evidence for the practical feasibility of using readily available higher-order theorem provers to support even very abstract and complex meta-logical investigations for IMLL within a categorical language suitably encoded in free logic embedded in HOL. 

\section{Conclusions}
\label{sec:conc}
This article has outlined an option for, and the potential of, formalizing category theory in higher-order logic to investigate meta-logical questions. The Isabelle/HOL proof assistant was used to formalize the notion of elementary topoi and to subsequently develop and formalize symmetrical monoidal closed categories expressing the denotational semantical model of intuitionistic multiplicative linear logic. This was carried out on the basis of (layered) shallow semantical embeddings, exploiting at the base layer a shallow semantical embedding of positive free higher-order logic in classical higher-order logic.

In addition to these meta-logical investigations, we outline with our work a possible way of building-up a category theory library in Isabelle to support reuse and application. We optimized our formalization to closely resemble mathematical notation and provided examples of how to use our formalized concepts. We hope that this will facilitate the future use of category theoretical notions in formalizations beyond just category theory itself.

\bibliographystyle{splncs04}
\bibliography{biblio_article}
%




\end{document}